\documentclass[a4paper, 11pt]{article} % 11pt font size and A4 paper size
\usepackage{graphicx}
\usepackage{caption}
\usepackage{amsmath}
\usepackage{yhmath}
\usepackage{hyperref}
\usepackage{rotating}
\usepackage{mathtools}
\usepackage{authblk}

\usepackage[
   margin=1in,
]{geometry}

\DeclareMathOperator*{\argmin}{arg\,min}

\title{A tomographic interpretation of structure-property relations for materials discovery}

\author[1,2]{Raul Ortega-Ochoa}
\author[3,4,5,6,7,8]{Al\'an Aspuru-Guzik}
\author[1,2]{Tejs Vegge}
\author[9]{Tonio Buonassisi}

\affil[1]{Department of Energy Conversion and Storage, Technical University of Denmark, DK 2800 Kgs. Lyngby, Denmark}
\affil[2]{CAPeX Pioneer Center for Accelerating P2X Materials Discovery, DK 2800 Kgs. Lyngby, Denmark}
\affil[3]{Chemical Physics Theory Group, Department of Chemistry, University of Toronto, 80 St. George St, Toronto, Ontario M5S 3H6, Canada.}
\affil[4]{Department of Computer Science, University of Toronto, 40 St. George St, Toronto, Ontario M5S 2E4, Canada.}
\affil[5]{Department of Chemical Engineering \& Applied Chemistry, 200 College St., University of Toronto, Ontario M5S 3E5, Canada}
\affil[6]{Department of Materials Science \& Engineering, 184 College St., University of Toronto, Ontario M5S 3E4, Canada}
\affil[7]{Vector Institute for Artificial Intelligence, 661 University Ave Suite 710, Toronto, Ontario M5G 1M1, Canada.}
\affil[8]{Acceleration Consortium, 700 University Ave, Toronto, Ontario M5G 1Z5}
\affil[9]{Department of Mechanical Engineering, Massachusetts Institute of Technology, Cambridge, MA 02139, USA}

\date{} % so that it doesnt print out the date

\begin{document}

\maketitle

\section*{Abstract}

Recent advancements in machine learning (ML) for materials have demonstrated that ``simple’' materials representations — \textit{e.g.}, the chemical formula alone without structural information — can sometimes achieve competitive property prediction performance in common-tasks. Our physics-based intuition would suggest that such representations are ``incomplete'', which indicates a gap in our understanding. This work proposes a \textit{tomographic interpretation} of structure-property relations of materials to bridge that gap by defining what is a material representation, material properties, the material and the relationships between these three concepts using ideas from information theory. We verify this framework performing an exhaustive comparison of property-augmented representations on a range of material's property prediction objectives, providing insight into how different properties can encode complementary information.

\section{Introduction}
Machine learning (ML) algorithms have been used as fast surrogate models to predict materials properties, a process often referred to as \textit{forward design}. More recently, ML has also been used to propose materials subject to design conditions in what has been similarly called \textit{inverse design} \cite{Zunger2018,doi:10.1126/science.aat2663}. Within the highly diverse ML approaches for materials discovery \cite{D4FD00153B}, there are two main innovation directions: The representation used (\textit{e.g}., chemical formula, crystallographic structure) and the model architectural design (\textit{e.g}., feed-forward, message-passing, transformer). Representation design focuses on providing a fair description of the material, while architecture is concerned with the flexibility of the space of functions that a model can approximate. A successful model requires the combination of both to capture  input-output (\textit{e.g}., structure-property) relations.

We think of the property $p$ of a material $M$ as the result of an unknown function $f(M) = p$ acting on the material. The general objective of a ML model is to tune a flexible and known, \textit{surrogate} function $f_{\theta}$ parametrized by $\theta$ so that it produces the same functional evaluations as the unknown \textit{true} function $f(M)$ on a set of $N$ samples $D=\{(M_{i}, p_{i})\}^{N}$ comprising the dataset\footnote{In practice we may include regularization terms, the $f_{\theta}(\cdot)$ might be an stochastic function and or we might take a probabilistic interpretation of the output of $f_{\theta}(\cdot)$, but the irreducible core of the learning problem could be best described with this equation.}. The task is then formulated as an optimization problem, \textit{i.e}., finding the set of parameters $\theta^{*}$ that minimize the difference between the true and the surrogate function on samples from the dataset. For a general cost or error function $\mathcal{L}$ measuring the difference of some ($f(M_{i}),f_{\theta}(M_{i})$) we write the solution of the optimization problem in Equation \ref{eq:old-general-formulation}.
\begin{equation}
    \theta^{*} = \argmin_{\theta} \ \frac{1}{N}\sum_{i=1}^{N}\mathcal{L}[f(M_{i}), f_{\theta}(M_{i})] = \argmin_{\theta} \ \frac{1}{N}\sum_{i=1}^{N}\mathcal{L}[p_{i}, f_{\theta}(M_{i})]
\label{eq:old-general-formulation}
\end{equation}

While multiple material representations exist in chemistry, physics, and in ML, we do not think of them as equally ``worthy'', which stems from the intuition developed from our physical understanding of the system in question. For example, human intuition would often suggest that the crystallographic structure is usually preferred over the chemical formula as the material representation when we aim to predict the heat of formation \cite{Davariashtiyani2023} or the band gap of a crystal because our training in physical science would suggest that we need to know the spatial structure to calculate or approximate the wave function of the material, from which then one can predict the band structure of the material.

Surprisingly, ML models often perform remarkably well despite using ``poor'' or ``incomplete'' material representations. This applies to many different classes of materials, such as predicting band gaps in 2D materials using heats of formation \cite{Dau2023}, the propensity for electrochemical intercalation of protons to predict Tc in superconductors \cite{Jiang2023}, the van der Waals radius for predicting the lattice thermal conductivity \cite{PhysRevMaterials.5.053801}, or deviation-combination descriptors in chemoinformatics and Quantitative Structure-Property Relations (QSAR) methods, like the python package MixtureMetrics for complex materials such as copolymer mixtures \cite{MAHINI2024101911}. Moreover, models based on the chemical formula — sometimes referred to as composition-based — lead on 6 out of 13 tasks on the Materials Benchmark \cite{Dunn2020} for general-purpose algorithms \cite{matbench-website}, and are often praised because they do not require the structure, which can be challenging to know in advance of a digital screening or inverse design campaign \cite{damewood2023representationsmaterialsmachinelearning, Wang2021, goodall2020predicting}. Finally, and more recently, large language models (LLMs) have been used to predict inorganic synthesizability from the chemical formula of a compound treated directly as text \cite{Kim2024}.

These counterintuitive results merit more in-depth analysis than they have received, since they exemplify an apparent fracture between our intuitive physical understanding of a given system or material and how ML processes information.

\subsection{Previous work}
 {\bf Tian et al. (2022)}\cite{tian2022informationnecessarysufficientpredict} compared the prediction accuracy of ML models with composition and composition-structure representations on the Material Project database \cite{10.1063/1.4812323}, observing similar performance for stable non-polymorphic materials and hypothesizing that composition embeds spatial information of ground-state structures in these materials.
 
 Among the most notable recent studies applying information theory to deep neural networks includes the work of {\bf Tishby et al. (2015)} \cite{7133169} introducing the information bottleneck and {\bf Shwartz-Ziv et al. (2017)} \cite{shwartzziv2017openingblackboxdeep} introducing the information plane to analyze training dynamics of neural networks.

\section{Methods}
\subsection{Problem formulation}
We must firstly state that a material representation is not the material. If we agree both crystallographic structure and the chemical formula are material representations then a representation must not be the material. Otherwise we would be forced to say that the crystallographic structure and the chemical formula are equivalent, which we know not to be true as there could be different spatial arrangements of the atoms in the chemical formula. Instead, we can think of a material representation as an approximation of the \textit{material essence} $R(M) \approx M$, where the material essence is an abstract and intangible object, one of a set of many possible materials $M \in \mathcal{M}$. Some representations are better than others based on the fidelity of the approximation, measured by the information content of the material preserved in its representation. In terms of information theory, we can state that $R_{a}(M)$ is a better representation than $R_{b}(M)$ if $I(R_{a}(M);M) > I(R_{b}(M);M)\ \forall M \in \mathcal{M}$, if it shares more information with the underlying material essence. Equivalently, the better material representation removes more ambiguity on the material given its representation, \textit{e.g.}, the crystallographic structure is a better representation than the chemical formula because it removes the degeneracy stemming from the polymorphs or structural isomers. 

Because a material representation is not the material, a dataset is not comprised of pairs ($M_{i}, p_{i}$), but ($R(M_{i}), p_{i}$), and we should correct Equation \ref{eq:old-general-formulation} as Equation \ref{eq:general-formulation}.
\begin{equation}
    \theta^{*} = \argmin_{\theta} \ \frac{1}{N}\sum_{i=1}^{N}\mathcal{L}[f(M_{i}), f_{\theta}(R(M_{i}))]
\label{eq:general-formulation}
\end{equation}

This subtle correction fundamentally changes the interpretation of the optimization problem. Its solution $f_{\theta^{*}}(\cdot)$ is a map between projections $f(M), R(M)$ of a common underlying material $M$. Notably, the line between what we call a property and what we call representations becomes blurry. In essence, they are projections of the same underlying material with the only distinction that what we often call a representation generally preserves more information about the material than a property, $I(R(M);M) > I(f(M),M)$ \footnote{A material representation tends to preserve more information of the material than a material property by design. Historically what we call a material representation aims to be a high-fidelity description in approximate one-to-one correspondence with the material. However, this is not generally the case for a property where there is often a one-to-many correspondence, indicating a loss of information.}. The threshold under which the representation-property line is drawn is not defined, and so we may refer to both simply as material \textit{projections}. Furthermore, if the line dividing representations and properties fades, then so does the distinction between forward and inverse design. If we insist on using this terminology, we can distinguish between them as follows: \textit{Forward design} finds the map from high to low-information projections of materials, whereas \textit{Inverse design} is the process of finding the map from low to high-information projections\footnote{Inverse design could then appear to violate the data processing inequality by creating information. This is not the case because for inverse design we generally do not predict the high-information projection directly but rather the probability distribution of the high-information projection from which we may sample. This output distribution reflects the uncertainty stemming from the loss of information.}.

\begin{figure}[ht]
    \centering
    \includegraphics[width=0.55\linewidth]{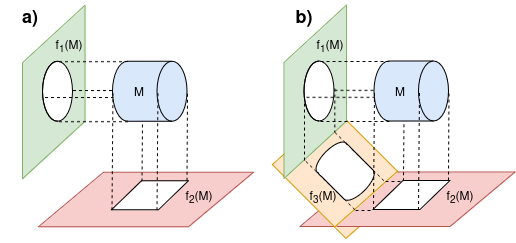}
    \caption{With enough shadows, we can uniquely identify the underlying object. Given the projections $f_{1}(M), \ f_{2}(M)$ the third $f_{3}(M)$ does not add extra information.}
    \label{fig:shadow_metaphor}
\end{figure}

As discussed, we may not have access to the material essence, but only to its projections. However, in the same way that given all possible (2D) shadows of a three-dimensional object, we can fully characterize it, if we have all the possible projections of the material essence then we have as much information as the essence itself. This process of reconstructing an object from its projections is reminiscent of the tomographic process, and so we name this interpretation of materials, properties, and 
representations the \textit{tomographic interpretation}. Note that there are many connections of this idea to the ideas of state tomography, \textit{e.g.}, the non-convex optimization challenge of reconstructing 3D particle structures 2D TEM (transmission electron microscopy) images in single-particle analysis (SPA) \cite{Lu2022} and process tomography in quantum information \cite{paris2004quantum} that we will explore in further work. The union of all possible projections would contain all the information about the material but it would be a highly redundant representation. Following with the shadow analogy, from Figure \ref{fig:shadow_metaphor}, given shadows $f_{1}(M), f_{2}(M)$ one has as much information of the three-dimensional object as if one adds a third shadow (Figure \ref{fig:shadow_metaphor}b) $f_{3}(M)$. So the number of projections needed to fully characterize the material is reducible to some amount.

It may now seem natural to define a \textit{minimally sufficient} representation of a material based on its projections. A \textit{sufficient} representation of a material is any of the possible representations that contains as much information as the material itself $I(R_{suf}(M);M) = I(M;M) , \ \forall  M \in \mathcal{M}$. A \textit{minimally sufficient} representation of a material may, however, be understood in two distinct ways: sufficient while minimal in the number of projections from a set of options\footnote{For example, for some particular application it may be useful to find the minimum subset of properties within those available that is sufficient for some task.}, or alternatively sufficient while minimal in information redundancy. Whether we use either of the definitions of minimal representation of a material, it is worth noting that there is nothing indicating such representation would be unique or factorizable into human-made descriptors\footnote{Furthermore, if we take the latter as definition of the minimal sufficient representation then we may call any such minimal sufficient representation the material essence, since there would be no possible way of making the distinction, because if there would be such way then the representation would not be sufficient.}.

This interesting discussion extends beyond the focus of the present work, and will be explored in future studies. We want to turn our attention back at Equation \ref{eq:general-formulation}: as discussed, the model's objective is to find the transformation between two projections of the same underlying object. From the data processing inequality, information can not be gained in post-processing $I(f(M),M) \leq I(M,M)$, so the sufficient information needed to find the map is, at worse, the material essence, but can be less. In other words, the minimal sufficient information to find a complete solution is task-dependent. This becomes more apparent in the shadow's prediction analogy of Figure \ref{fig:shadow_metaphor}: in the same way we do not need to know the color of the three-dimensional object to predict its shadow, there might be information of the material essence we do not need to account for to solve the optimization problem in Equation \ref{eq:general-formulation}.

Finally we must discuss the role of the dataset. Sufficiency and minimality have been defined for materials within all that are possible. In practice, datasets are hardly representative of all possible materials but rather a subset thereof (\textit{e.g.}, perovskites \cite{C1EE02717D}) which relaxes the requirements for sufficiency and minimality from the general case into the particular case of the dataset.

We may now revisit the central question of our work: How can ``simple'' representations used in ML for materials obtain ``unreasonably'' good performance? Under the lens of the developed concepts in this work, we realize that the surprise permeating this question is finding that we needed less information to perform the task than we initially thought. Regarding the specific dilemma between chemical formula vs structure representation in ML for materials, we propose a causal explanation on why composition may be sufficient: Generally, the elemental composition and stoichiometry constrain the structure of a material, because there are limited amount of spatial arrangements (polymorphs) for a given composition and stoichiometry due to atom's valence rules. In particular,  datasets may not include all of the possible polymorphs. Then, if datasets particularly limit the generally constrained number of polymorphs, there could be a near one-to-one correspondence between chemical formula and structure. Therefore, even if the structure would contain more useful information to perform the task, this degeneracy is not reflected in the dataset, and the optimization task formally described by Equation \ref{eq:general-formulation} can be solved without accounting for explicit materials geometry.

\begin{figure}[ht]
    \centering
    \includegraphics[width=0.35\linewidth]{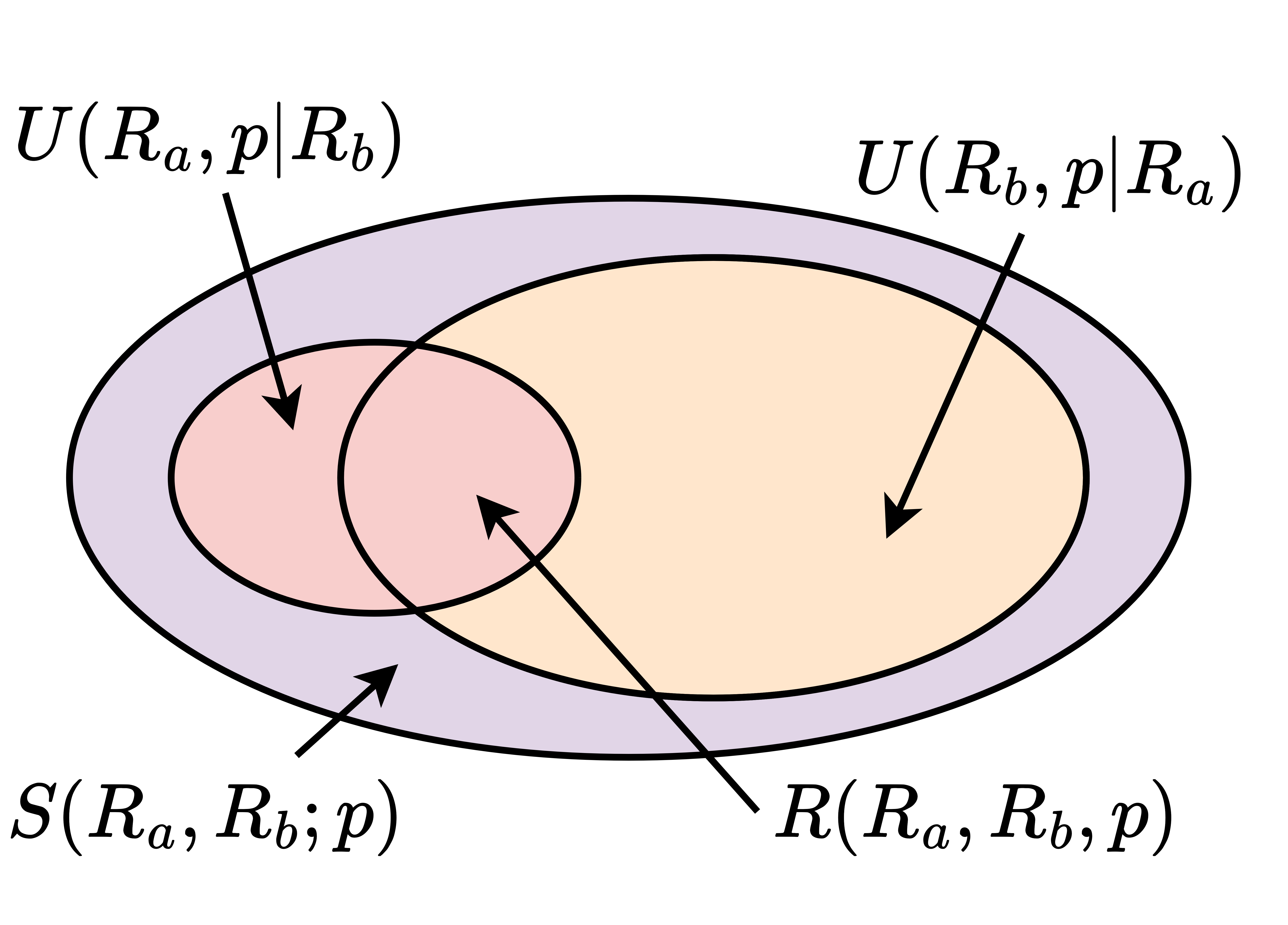}
    \caption{PID diagram of the unique, synergetic and redundant information contributions of two source variables $R_{a}, R_{b}$ and a target variable $p$. Adapted from \cite{WIBRAL201725}.}
    \label{fig:pid-diagram}
\end{figure}

As discussed, different projections can encode different information of the material, and so combining them may gather more information than any of its individual projections\footnote{Note that the combination of projections is itself a projection.}. The benefit of including an extra projection to augment a representation depends on how this new source of information interacts with that already present.
To study the information interactions between multiple variables, information theory has been extended in different ways \cite{5392532,garner1962uncertainty,Studeny1998,Tononi1994-pm,McGill1954,SUNHAN198026,ICA03:Bell,Gawne2758}; one of such ways is partial information decomposition (PID)\cite{williams2010nonnegativedecompositionmultivariateinformation}, which has the advantage that it provides insight into how dependencies are distributed among the multiple variables and the components are non-negative. PID decomposes the information of some target variable captured by multiple sources into a set of non-negative terms: The unique, redundant and synergetic contribution. Using the PID framework, we factorize the information encoded about some target variable $p$ by two other variables $R_{a}, R_{b}$ in Equation \ref{eq:pid}. These four contributions can also be visualized in the PID venn diagram in Figure \ref{fig:pid-diagram}.

\begin{equation}
    I(R_{a},R_{b};p) = \underbrace{U(R_{a},p|R_{b})}_{\text{Unique $R_{a}$}} + \underbrace{U(R_{b},p|R_{a})}_{\text{Unique $R_{b}$}} + \overbrace{S(R_{a},R_{b};p)}^{\text{Synergetic $R_{a}, R_{b}$}}
    + \underbrace{R(R_{a},R_{b},p)}_{\text{Redundant $R_{a}, R_{b}$}}
    \label{eq:pid}
\end{equation}

Under the light of the tomographic interpretation and PID some surprising results prevalent in the literature become plainly clear. Perhaps one particularly representative example is \cite{D4FD00113C}: \textit{``We’ve hypothesized that the property-to-structure mapping becomes unique when a sufficient number of properties are supplied to the models during training. This hypothesis has several important corollaries if true. It would imply that data-scarce properties can be completely determined using a set of more accessible molecular properties.''} From the standpoint of the tomographic interpretation one can also justify the use of extra properties to augment representations, for example, the elemental property matrix in FTCP \cite{REN2022314} includes electronegativity and covalent radius into the atomic features, which was justified empirically because its addition resulted in better performance. Or similarly \cite{Onwuli_Butler_Walsh_2024} extend atomic embedding vectors with oxidation states and find that ionic representations improved performance on tasks linked to the electronic structure of a material. Perhaps more illuminating, after the introduction of equivariance into machine-learning potentials, a substantial part of the research has focused on enhancing the representation, for example by including long-range electrostatics \cite{Shaidu2024}, magnetic moments (CHCnet \cite{deng2023chgnetpretraineduniversalneural}), or spin states (Spookynet \cite{Unke2021}).

A consistent finding across multiple studies extending representations through the incorporation of properties is the observation of tasks or materials, where this process leads to performance improvements in certain cases, while in others, performance remains unaffected. Proposing a causal explanation in terms of human-based descriptors for these observations would need to examine every case's particularities. However, under the lens of the tomographic interpretation with PID, we can provide an explanation on \textit{why} that could be the case, even if we may not know generally the \textit{how} because the mechanisms are case-specific.

For some property (in this context also referred in the literature as feature) $R_{a}$ to be added on top of an already existing representation $R_{b}$ and improve performance on some task $p$ (the target property) it needs to encode information about the task that is not found on the original representation: the first, unique term in Equation \ref{eq:pid}, which gives rise to the third, synergetic term. \textit{E.g.}, Because including the oxidation states of the atomic elements improved performance on certain tasks \cite{Onwuli_Butler_Walsh_2024} it must be this information is relevant and was not originally available in the original representation.

\subsection{Verification Experiments}

The objective of the following experiments is to qualitatively estimate what type of information is encoded in two standard material representations: chemical formula — or composition — and composition-structure. This, in turn, can shed light on how may we augment this representations depending on the objective task. To this end we use a 2020 snapshot of Materials Project \cite{10.1063/1.4812323} containing 126 325 materials\footnote{We note that for 74 635 (59.08\%) database entries there is a one-to-one correspondence between material identifier and chemical formula, which was hypothesized to be part of the cause why chemical formula could almost suffice as a representation for some datasets.}. There are different approaches of estimating the mutual information between random variables \cite{PhysRevE.69.066138,10.1063/5.0217960} however these methods do not scale to large number of dimensions, which is the natural regime of ML. Therefore we aim to indirectly assess this quantity through the feature importance. The feature importance of some variable for some task is here defined as the relative change in the performance of the augmented representation with respect to the original representation. And we use the mean absolute error (MAE) in the test set as a measure of performance. In terms of feature importance, adding some new property/feature encoding non-redundant information needed for predicting some target will result in a decrease in the error with respect to the non-augmented baseline, and in the case it does not encode relevant information, or the information was already accounted for, then there will be no change in the error with respect to the non-augmented baseline, because the performance can not be harmed by including more information since mutual information is non-negative, and strictly zero if the variables are independent\footnote{We must acknowledge that in practice including some extra property could harm the performance, not as a result of the information it conveys — which is the subject of discussion here — but, for example, if it is poorly scaled so the variance is so large it perturbs the learning process.}.

While most of the discussion in this work has focused on the information content of different representations, in practice different model architectures are not equally capable of extracting information from representations and the effect we would like to attribute to the change in representation may well be due to the architecture. In order to compare representations we need to use the same architecture, and so to compare results from composition-restricted and composition-structure representations CGCNN \cite{PhysRevLett.120.145301} was modified so the structure component may be omitted if requested. For CGCNN the structure component can be omitted by ``fooling'' the bond perception algorithm used to initialize the CGCNN graph so that every site is bonded to each other at equal distance.

\begin{figure}[ht]
    \centering
    \includegraphics[width=\linewidth]{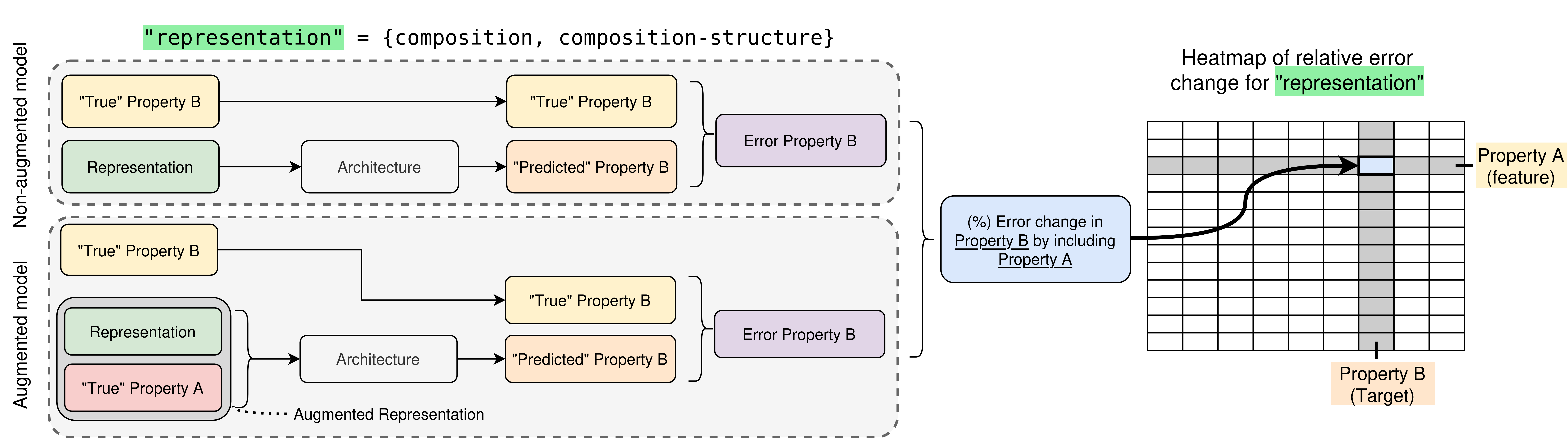}
    \caption{Schematic process of calculating feature importance for ``Property A'' in the task of predicting ``Property B'' and how these results are summarized in the heatmaps of Figures \ref{fig:heatmap-3d}, \ref{fig:heatmap-comp}. This process involves training two models (non-augmented and augmented), repeated five times for 13 different ``Property A'', for 9 ``Property B'', for a composition and composition-structure representation.}
    \label{fig:schema-experiments}
\end{figure}

The feature importance calculated is sensible to the training-validation-testing split and network initialization, and so to obtain a more robust estimation we repeat each experiment for five different seeds with a constant 60-20-20 split ratio and aggregate the results in the mean, maximum, and minimum feature importance. Every model was trained with the original hyperparameters and for 500 epochs from which the best model is chosen based on the error in the validation set. The feature importance was computed for predicting nine properties: band-gap, density, energy above hull, energy, energy per atom, formation-energy per atom, number of sites, total-magnetization and volume. Each experiment uses a different property/feature out of thirteen: including the nine already mentioned, is-hubbard, oxide-type, spacegroup-crystal-system and spacegroup number. This process is repeated for five different dataset splits for both the composition-based and composition-structure representations, resulting in a total of 2340 models trained\footnote{Total of 2 separate experiments (composition and composition-structure) for 9 target properties, with 13 input properties, repeated for 5 seeds and 2 experiments (baseline and augmented representation) needed to compute the feature importance.}. The schema in Figure \ref{fig:schema-experiments} represents the steps required in computing the feature importance of some ``Property A'' for predicting ``Property B'', and how the results of this process are organized in the heatmaps in Figures \ref{fig:heatmap-3d}, \ref{fig:heatmap-comp}.

\section{Results}

\begin{figure}[ht]
    \centering
    \includegraphics[width=\linewidth]{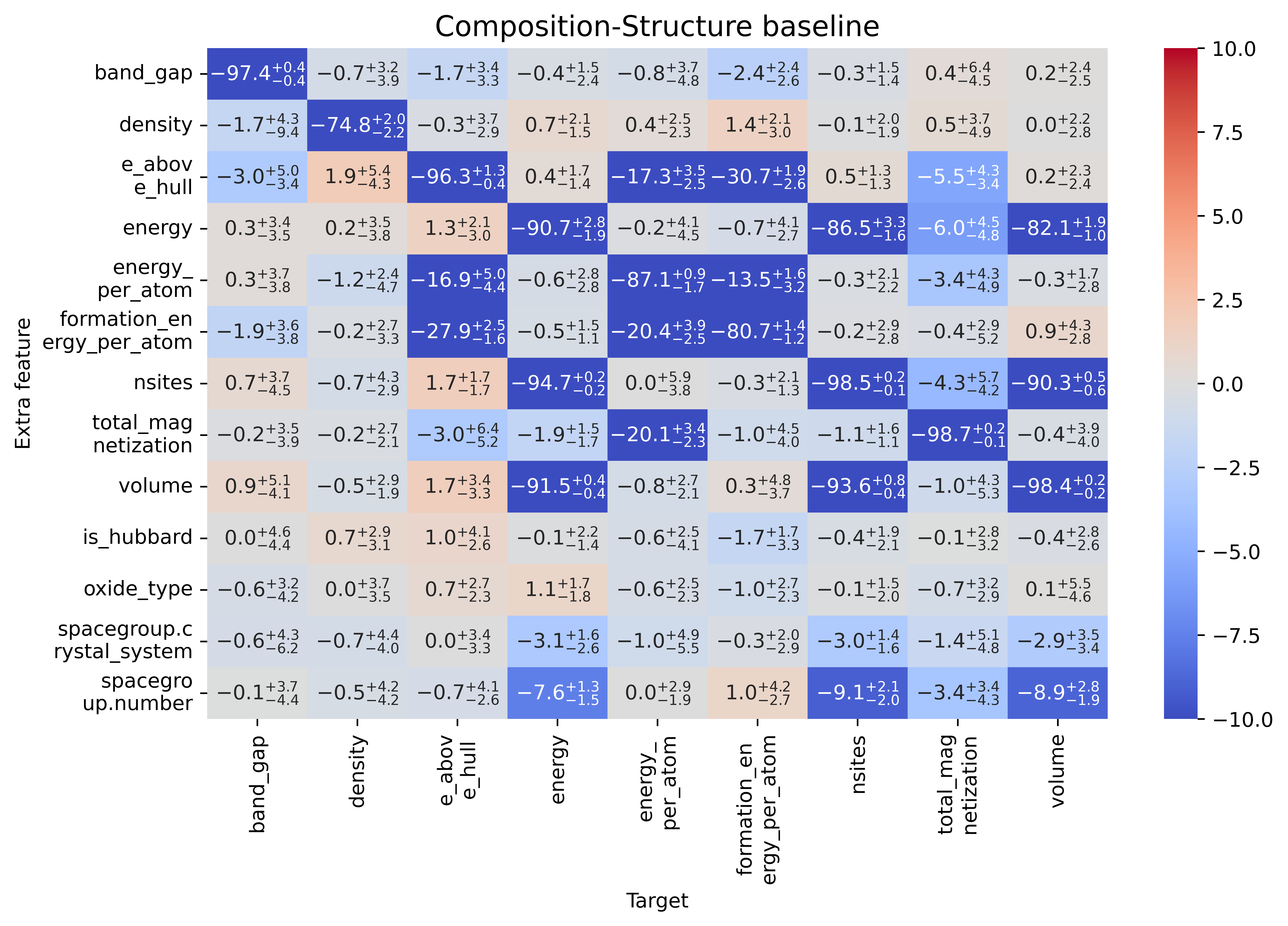}
    \caption{Percentage change in MAE of an augmented vs a non-augmented composition-structure representation.}
    \label{fig:heatmap-3d}
\end{figure}

\begin{figure}[ht]
    \centering
    \includegraphics[width=\linewidth]{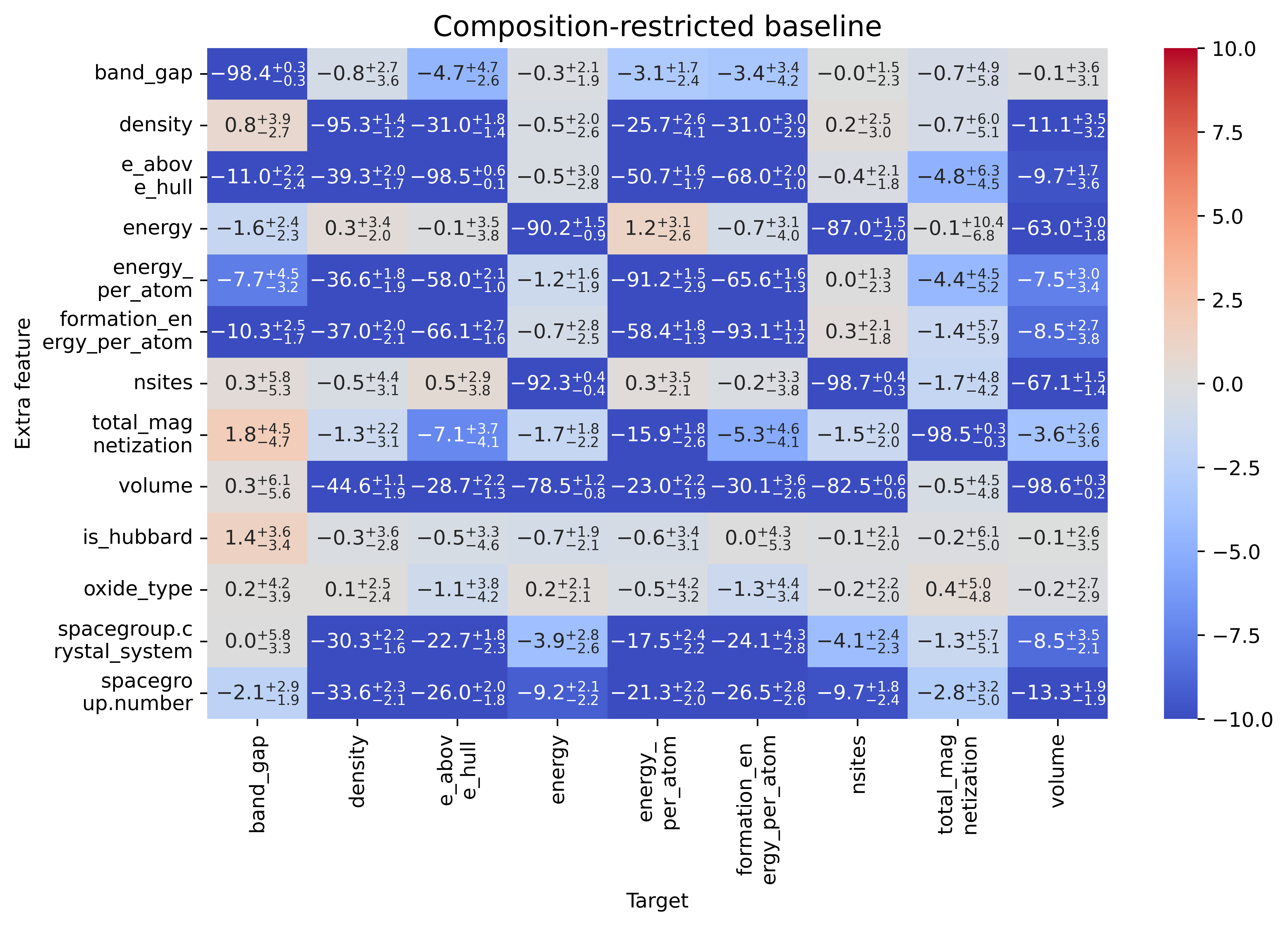}
    \caption{Percentage change in MAE of an augmented vs a non-augmented composition-restricted representation.}
    \label{fig:heatmap-comp}
\end{figure}

The heatmaps in Figures \ref{fig:heatmap-3d}, \ref{fig:heatmap-comp} illustrate the result of introducing different properties (\textit{y}-axis) to augment the representation for a range of tasks (\textit{x}-axis), using respectively a composition-structure (Figure \ref{fig:heatmap-3d}) or composition-restricted (Figure \ref{fig:heatmap-comp}) representation as baseline. A negative value, in blue, corresponds to a relative reduction of the prediction error by introducing some extra property. Note that the focus of this analysis is the relative change in performance and not the performance itself, so a significant improvement in performance does not mean the model is accurate, but rather that however accurate it is better after introducing the property. We can use the first row, sixth column entry in Figure \ref{fig:heatmap-3d} as an example on how to read these figures: this entry corresponds to the effect of using band-gap to augment the representation for predicting formation energy per atom, which was shown to reduce the error by $-2.4^{+2.4}_{-2.6}\%$. The error change is compatible, in its upper bound, with zero indicating that the worst-performing augmented representation performed on par with the best-performing non-augmented. In this study, we will consider the addition of a property to have a meaningful impact on error reduction only if the reduction is statistically significant, such that its confidence interval (min-max range) does not overlap with zero\footnote{We choose the min-max confidence interval instead of the standard deviation as a more conservative estimate of the true range, since the standard deviation could not be accurately estimated given the relatively scarce (5) number of repetitions.}.

Both the composition-structure and composition-restricted experiments in Figures \ref{fig:heatmap-3d}, \ref{fig:heatmap-comp} exhibit some common overall patterns. Firstly, they show an intense blue diagonal line corresponding to the case where the target is used as a feature, in which case the problem is trivial. Secondly, for all entries whose mean performance change is positive, shown in red (which indicates worse performance of the model after introducing the property), the error bounds overlap with 0\% and are thus not statistically significant in accordance with our criteria. Therefore, these results are in line with the statement that more information, even if independent of the target, does not harm performance. Thirdly, and more interestingly, in both experiments the heatmaps are note entirely symmetric: \textit{e.g.}, in Figure \ref{fig:heatmap-3d} including total-magnetization reduces the energy-per-atom prediction error  $-20.1^{+3.4}_{-2.3}\%$ whereas energy-per-atom reduces the total-magnetization prediction error $-3.4^{+4.3}_{-4.9}\%$, compatible with 0\% in its upper bound. If the difference in relative change was small, we may be inclined to attribute this to errors or noise during training, but in this particular example the gap is so large that it demands some other, perhaps complementary, explanation: unlike mutual information, partial information is not symmetric \cite{williams2010nonnegativedecompositionmultivariateinformation} \textit{e.g.}, it could be that energy-per-atom and total-magnetization share information, however it may be that energy-per-atom is well captured by the original representation and total-magnetization is not. Therefore including total-magnetization to predict energy-per-atom adds more non-redundant information than including energy-per-atom to predict total-magnetization.

Comparing the overall results between the composition-restricted (Figure \ref{fig:heatmap-comp}) vs composition-structure (Figure \ref{fig:heatmap-3d}) we can clearly see for composition-restricted including properties generally helped more than for composition-structure, which suggests that these properties encode valuable information for the target properties (or they would not decrease the error), and that these properties are better captured in the composition-structure than simply in the composition (the relative improvement is lesser in composition-structure for the same property and target). In order to inspect individually which properties helped more for which others in Table \ref{tab:heatmaps} we include for each experiment (Figure \ref{fig:heatmap-3d} in column ``Structure'', Figure \ref{fig:heatmap-comp} in column ``Composition''), for each target property (rows in column ``Target'') which are the properties that, using the significance criteria introduced, had an impact on the reduction of prediction error. Then, in column ``In Struct. but not Comp'' we list the elements of composition column not in the structure column, which can be interpreted as properties that are exclusively embedded in the the representation when the structure is included. Counting the frequency of each of the different properties in the last column we can see that, in order, the most frequent properties embedded only when we include structure into the representation is firstly the spacegroup crystal system, followed then by spacegroup-number, density and volume. Notably, these are most explicitly structure-dependent and thus this result is consistent with our physical intuition.

\begin{table}[ht]
\begin{tabular}{|l|l|l|l|}
\hline
 Target & Structure & Composition & In Struct. but not Comp.  \\ \hline\hline
 band-gap & \{\} & \{2,4,5\} & \{2,4,5\}   \\ \hline
 density & \{\} & \{2,4,5,8,11,12\} & \{2,4,5,8,11,12\}   \\ \hline
 e-above-hull & \{4,5\} & \{1,4,5,7,8,11,12\} & \{1,7,8,11,12\}  \\ \hline
 energy & \{6,7,8,11,12\} & \{6,8,11,12\} & \{\}  \\ \hline
 e-per-atom & \{2,5,7\} & \{0,1,2,5,7,8,11,12\} & \{0,1,8,11,12\}  \\ \hline
 formation-e-per-atom & \{2,4\} & \{1,2,4,7,8,11,12\} & \{1,7,8,11,12\}  \\ \hline
 number-sites & \{3,8,11,12\} & \{3,8,11,12\} & \{\}  \\ \hline
 total-magnetization & \{2,3\}  & \{\} & \{\}   \\ \hline
 volume & \{3,6,12\} & \{1,2,3,4,5,6,7,11,12\} & \{1,2,4,5,7,11\}  \\ \hline
\end{tabular}
\caption{Properties improving predictive performance for different target properties. Indexes of the properties: 0: band-gap, 1: density, 2: e-above-hull, 3: energy, 4: e-per-atom, 5: formation-e-per-atom, 6: number-sites, 7: total-magnetization, 8: volume, 9: is-hubbard, 10: oxide-type, 11: spacegroup-crystal-system, 12: spacegroup-number.}
\label{tab:heatmaps}
\end{table}

\section{Conclusion}

In this work we introduce a conceptual framework for ML in materials discovery in an effort to explain the apparent conflict existing between our physical intuition of what constitutes a ``good'' representation and how ML models achieve competitive performance with ``simple'' representations in a range of tasks. To achieve this, we provide a series of definitions outlining what constitutes a material representation, a material property, and the material itself, as well as how these three concepts are interconnected. We propose an interpretation of the material as a separate, abstract and inaccessible object from which we can only observe its representations and properties, which we unify under the concept of material \textit{projections}. We discuss how this framework sheds light in yet unexplained and ``surprising'' commonly observed results, including the remarkable performance of ``simple'' representations, and how ML representations can be augmented by including other related properties to improve predictive performance for some tasks.

We verify our framework by conducting a comprehensive comparison between property-augmented and non-augmented representations, across both composition-restricted and composition with structure representations, and for various property prediction tasks. From these experiments, we gain valuable insight into how different material \textit{projections} encode complementary information. Finally, we show that explicitly structure-dependent properties like spacegroup-derived properties, the density or the volume, are more widely captured in the composition-structure representation, in accordance with our physical intuition.

Our proposed framework, which we call the tomographic interpretation of structure-property relations of materials, opens the door to a series of interesting corollaries which we have only began to explore. For example, multi-property inverse design should achieve better results the more properties one conditions on, given that every projection can encode complementary information to aid the reconstruction.

The tomographic interpretation motivates a change in perspective of ML for materials discovery and sets the stage for further research on materials representations.

\section{Acknowledgments}
The authors acknowledge financial support from the Technical University of Denmark (DTU) through the Alliance Ph.D. scholarship, from the Independent Research Foundation Denmark (0217-00326B), the Pioneer Center for Accelerating P2X Materials Discovery (CAPeX), DNRF grant number P3, the Novo Nordisk Foundation Data Science Research Infrastructure 2022 Grant: A high-performance computing infrastructure for data-driven research on sustainable energy materials, Grant no. NNF22OC0078009. This research was undertaken thanks in part to funding provided to the University of Toronto’s Acceleration Consortium from the Canada First Research Excellence Fund CFREF-2022-00042. Computational resources used in preparing this research were provided by the Acceleration Consortium. A.A.-G. thanks Anders G. Frøseth for his generous support. A.A.-G. also acknowledges the generous support of the Canada 150 Research Chairs program. A.A.-G. and T.B. are CIFAR Fellows in the Accelerated Decarbonization Program; this research is based in part on work supported by CIFAR through a catalyst award.

%\bibliographystyle{unsrt}
%\bibliography{bibliography.bib}

\end{document}